\documentclass[iop]{emulateapj}
\usepackage{epstopdf}
\usepackage{amsmath, amsthm, amssymb}
\usepackage{graphics, subfigure}
\usepackage{fancybox}
\usepackage{graphicx}
\usepackage{color}
\usepackage{epsfig}
\usepackage{grffile}

\newcommand{\teff}{T_{\rm eff}}

\newcommand{\sn}{{\mathrm S}/{\mathrm N}}
\newcommand{\snth}{{({\mathrm S}/{\mathrm N})}_{\mathrm thres}}

\newcommand{\days}{\,{\rm days}}
\newcommand{\da}{\,{\rm d}}
\newcommand{\kepler}{{\it Kepler}\, }
\newcommand{\tdur}{{t_{\rm dur}}}

\newcommand{\ntr}{N_{\rm tra}}

\newcommand{\thres}{{\rm thres}}
\newcommand{\eff}{{\rm eff}}
\newcommand{\gyr}{{\rm Gyr}}
\begin{document}
\title{
Fast Rise of ``Neptune-Size'' Planets ($4-8 R_\earth$) from $P\sim10$ to $\sim250$ days \newline
-- Statistics of {\it Kepler} Planet Candidates Up to 
$\sim 0.75 \rm AU$}
\author{Subo Dong\altaffilmark{1,2} and Zhaohuan Zhu\altaffilmark{3,4}}
\altaffiltext{1}{Institute for Advanced Study, 1 Einstein Dr.,
Princeton, NJ 08540, USA}
\altaffiltext{2}{Current Address: Kavli Institute for Astronomy and Astrophysics, Peking University, Yi He Yuan Road 5, Hai Dian District, Beijing, 100871, China}
\altaffiltext{3}{Department of Astrophysical Sciences, Princeton
University, Princeton, NJ, 08544}
\altaffiltext{4}{Hubble Fellow}
\begin{abstract}

We infer the period ($P$) and size ($R_p$) distribution 
of \kepler transiting planet candidates 
with $R_p\ge 1 R_\earth$ and $P<250 \da$ hosted by 
solar-type stars. 
The planet detection efficiency is computed by using measured 
noise and the observed timespans of the light curves for 
$\sim 120,000$ \kepler target stars. We focus on deriving the 
shape of planet period and radius 
distribution functions. We find that for orbital period $P>10\da$, 
the planet frequency d$N_p$/d$\log$P
for ``Neptune-size'' planets ($R_p = 4-8 R_\earth$) increases 
with period as $\propto P^{0.7\pm0.1}$. In contrast, 
d$N_p$/d$\log$P for ``super-Earth-size'' ($2-4 R_\earth$) as well 
as ``Earth-size'' ($1-2 R_\earth$) planets are consistent with a nearly flat distribution as a function of period ($\propto P^{0.11\pm0.05}$
and $\propto P^{-0.10\pm0.12}$, respectively), 
and the normalizations are remarkably similar (within a factor of 
$\sim 1.5$ at $50 \da$). Planet size distribution evolves with period, and generally the relative fractions for big planets ($\sim 3-10 R_\earth$) increase with period. The shape of the distribution function is not 
sensitive to changes in selection criteria of the sample.
The implied nearly flat or rising planet frequency at 
long period appears to be in tension with the sharp decline at $\sim 100\da$
in planet frequency for low mass planets (planet mass $m_p < 30 M_\earth$) 
recently suggested by HARPS survey. Within $250 \days$, the cumulative frequencies for Earth-size and super-Earth-size planets are remarkably similar ($\sim 28 \%$ and $25\%$), while Neptune-size and Jupiter-size planets are $\sim 7\%$, and $\sim 3\%$, respectively.  A major potential uncertainty
arises from the unphysical impact parameter distribution of the 
candidates. 
\end{abstract}

\section{Introduction}
The {\it Kepler} mission provides an unprecedented opportunity to study 
the size and period distribution of extrasolar planets down to 
Earth radii within yr-long orbits by making high-precision 
($\sim 10^{-4}$), high-cadence ($\sim 30 \min$) and nearly-continuous 
monitoring of $\sim 10^5$ stars over years. Based on 
the transiting planet candidates discovered from the first 4 months       
of {\it Kepler} data (\citealt{kep4}, hereafter B11), 
\citet{howard} (hereafter H12) made a statistical inference of 
the frequency for planets with radii $R_p\ge2 R_\earth$. H12 
found that planet frequency increases for decreasing radii, and that it 
drops sharply for planets with very close-in orbits ($P<10\da$). They 
claimed that the \kepler planet frequencies are consistent with 
those found by radial-velocity (RV) surveys \citep{mayor09,howard10}.
Several other studies also use the B11 sample to study planet distribution.
\citet{break} found that there is a break in the radius distribution
of B11 candidates at $\sim 3 R_{\earth}$. By extrapolating 
the detection efficiency deduced by H12 and applying a 
maximum-likelihood approach, \citet{youdin} fitted the distribution 
of B11 candidates down to $0.5 R_\earth$ and found a relative 
deficiency of $\sim 3 R_{\earth}$ planets at $P<7 \days$. 
\citet{shao} and \citet{etaearth} attempted 
to extrapolate the planet frequency obtained from B11 candidates to 
estimate the fraction of Sun-like stars that host habitable 
Earth-like planets.

The latest release of {\it Kepler} based on 16 months (quarters 
Q1-Q6) of data (\citealt{kep16}, hereafter B12) has 
increased the number of known planet candidates by a factor of 
$\sim 2$ (from $\sim 1200$ to $\sim 2300$). As expected, there is 
a large gain in planet candidates at long periods as well as 
small radii compared to B11. According to 
B12, there is also a considerable unexpected gain relative to B11 for 
short-period planets merely due to the effects of increasing the 
length of the observing windows, and the implied lower-than-expected 
efficiency of the planet search pipeline employed by B11 may affect 
the above-mentioned statistical results. One important improvement 
in B12 is that for the first time the {\it Kepler} team stitched  
different quarters together in the transit search, which particularly  
increased the robustness of the search for long-period planets. In fact, 
two independent automatic planet searches on Q1-Q6 data by 
\citet{huang} and \citet{corot} as well as crowd-sourced human 
identifications by Planet Hunters \citep{planethunter} only identified  
a total of $\sim 10\%$ more new planet candidates than those found 
by B12, suggesting that the B12 searches are likely highly efficient.

We derive planet frequency as a function of period and planet radius  
using {\it Kepler} planet candidates discovered by B12 as well
as those found by other groups. Like the majority of the 
works on \kepler statistics to date, we do not distinguish 
planet candidates from planets, i.e., we assume a low false positive 
rate (see \S~\ref{sec:disc} for further discussion).
The transit planet detection 
efficiency is calculated for each \kepler star using the measured 
photometric noise of its light curve and the observed timespan 
(excluding gaps and missing quarters). In addition, the 
geometric bias for circular orbits is taken into account. 
We focus on determining the relative
frequency for planets with various radii as a function of period for 
Sun-like hosts. We find that the distribution of reported 
impact parameters is unphysical, potentially posing a major
 uncertainty in the overall normalization of the planet distribution 
 function.
We do not distinguish planets in the single or 
multiple transit systems to derive the planet multiplicity function 
\citep{tremainedong} and we also ignore any possible bias in the detection of single and multiple systems. 
The \kepler planet frequency derived below extends
down to $R_p \ge 1 R_\earth$ with period up to $\sim 250\days$.
This can be compared with the planet frequency inferred from 
RV searches by 8-yr HARPS survey, which is sensitive
to long-period ($P\gtrsim 100\days$) super-Earth and Neptunes with
masses $M_p<30 M_\earth$ \citep{mayor11}.

\section{Issues with selecting the {\it Kepler} star and planet sample}

\subsection{The {\it Kepler} Input Catalog}
\label{sec:kic}
Planet frequency is usually defined with respect to an ensemble 
of host stars that share similar physical properties. Stellar 
type, metallicity, age and population may have impacts on the 
frequency of planets. The {\it Kepler} target stars were selected 
based on multi-band photometry (documented in the Kepler Input Catalog, KIC), and the selection was focused
on finding solar-type stars to search for Earth analogues. 

The KIC photometry is
most sensitive to the effective temperature $T_{\eff}$, which is 
less reliable for constraining surface gravity $\log g$ (particularly  
unreliable for cool stars) and has little sensitivity to metallicity. 
We do not attempt to study the planet frequency as a function of 
metallicity, which would require comprehensive spectroscopic 
follow-up. The relatively large uncertainty in $\log g$ may 
have a serious impact on the study of frequency. Unreliable $\log g$ 
estimates may introduce ambiguity between dwarfs and 
sub-giants/giants with the same $T_{\eff}$. Furthermore, errors in 
$\log g$ dominate the uncertainties in 
the stellar radius measurement, which translates into uncertainty 
in the planet radius since only planet-to-star radius ratios
are measured from transit light curves.

\begin{figure*}
\includegraphics[scale=0.9]{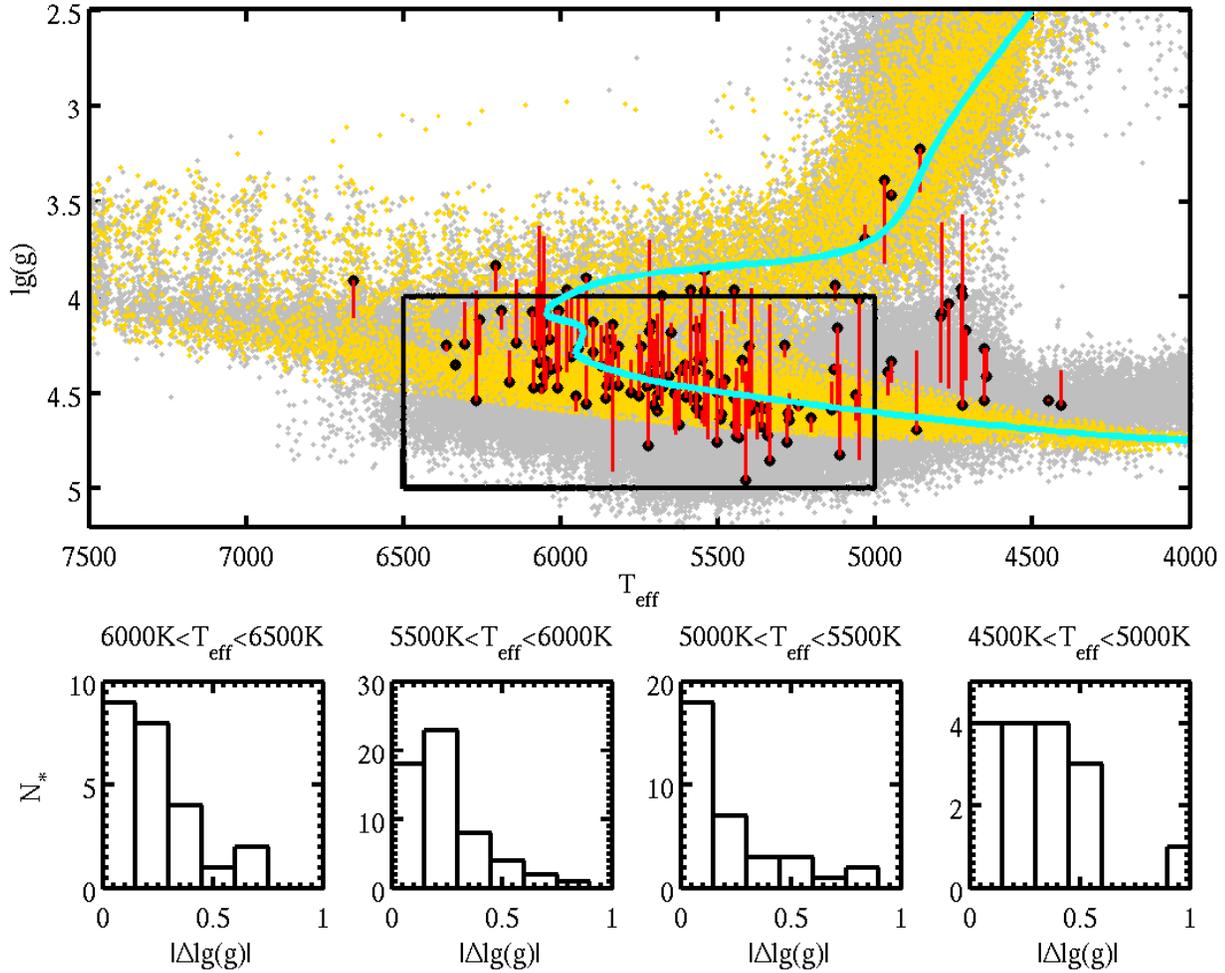}
\caption{Upper panel: $T_{\rm eff}$ and log $g$ from  
KIC catalog for all {\it Kepler} target stars (gray dots) as well as
the ``corrected'' stellar parameters derived by matching 
Yonsei-Yale isochrones following the approach in B12 
(yellow dots) [see \S~\ref{sec:kic} for detailed discussion]. 
The cyan line is the Yonsei-Yale isochrone for solar age at 
solar-metallicity. We also 
highlight 104 stars with 
accurate stellar parameters derived from high-resolution spectroscopic follow-ups
from \cite{kepmet}. The $T_{\eff}$ and log $g$ for these stars from KIC are plotted
as the black solid dots, while the log $g$ value from the spectroscopic measurements
are plotted at the end of the red lines connecting from the KIC values.
Lower panels: we divide the104 stars into four different 
temperature bins and calculate the difference between
KIC and spectroscopically measured log $g$ values. The average dispersion is about 0.3dex, except for the lowest temperature bin ($4500-5000 K$) and we exclude stars with $T_{\rm eff} < 5000 K$ in our stellar sample. The selected stellar sample is within 
the black box based on the ``corrected'' parameters 
shown in the upper panel.
\label{fig:kic}}
\end{figure*}

To study the uncertainty in $\log g$, we use the high-precision
stellar parameters derived from high-resolution spectroscopic 
follow-up of more than a hundred \kepler planet host stars 
by \citet{kepmet}. In the upper panel of 
Fig.\ref{fig:kic}, the KIC 
$T_\eff$ and $\log g$ are plotted as black solid dots, and 
the $\log g$ values from 104 spectroscopic measurements are 
plotted at the end of the red lines connected to the KIC values.
The majority of the stars in the KIC have $T_{\eff}$ between $4500K$
and $6500K$, and we divide these stars into four equal bins in temperature. For each bin, the average difference 
between the two sets of measurements $\Delta{\log g}$ is 
$\pm 0.1$, with no strong systematic preference in sign. The 
average dispersion is about $0.3$dex, except for the bin with 
$4500K<\teff<5000K$, which has a dispersion of $0.4$dex. In 
the lower panel, the histogram of $|\Delta{\log g}|$ is shown;  
for the bin with $4500K<\teff<5000K$, $50 \%$ of the stars 
have $|\Delta{\log g}|>0.3$dex, while $\lesssim 30\%$ stars 
have $|\Delta{\log g}|>0.3$ for the three other bins. It seems 
that the problem with $\log g$ uncertainty is most severe for stars 
with $\teff<5000K$, and we choose not to include them in our 
stellar sample for this study. The averaged dispersion in 
$\log g$ for the chosen stellar sample is therefore $0.3$dex, 
which translates into $0.15$dex dispersion in the planet
radius estimate.

B12 noted that a considerable fraction of KIC stellar 
parameters were not consistent with known stellar physics. 
They matched the KIC $\teff$, $\log g$ and [Fe/H] with 
Yonsei-Yale isochrones \citep{yaleyonsei} by minimizing
$(\delta \teff/200K)^2+(\delta \log g/0.3)^2 + 
(\delta {\rm [Fe/H]} / 0.4)^2$, where $\delta$ is the 
difference in the KIC and Yonsei-Yale parameters. They reported
the stellar parameters (and the derived planet parameters) using
the ``corrected'' values from Yonsei-Yale. We note that the 
``corrected'' stellar parameters do not match the spectroscopic
measurements from \citet{kepmet} better than those from the KIC. Nevertheless,
they are at least self-consistent for each star according to 
the known laws of stellar physics (e.g., the parameters match 
the theoretical mass-radius relation).
We follow the procedure by B12 and adopt ``corrected'' 
parameters throughout this paper. In Fig.\ref{fig:kic}, the 
``corrected'' $\teff$ and $\log g$ are plotted as yellow dots
and the KIC values are shown as gray dots (the Yonsei-Yale 
isochrone for $5 \gyr$ with solar 
metallicity is shown in cyan). It is interesting to note that 
many stars at $4500K <\teff < 5000K$ have KIC $\log g$ values
inconsistent with any reasonable isochrones.

Our stellar sample consists of \kepler stars with 
$5000K<\teff<6500K$ (approximately corresponding to
K2-F5 dwarfs) and $4.0<\log g<5.0$. These limits are 
shown as a black box in Fig.\ref{fig:kic}. We also exclude stars with \kepler magnitude $m_K>16$, which consist of a 
negligible fraction of \kepler stars and have little sensitivity
to planets. The sample includes a total number of stars of 
$N_* = 122328$ . 

\subsection{Impact Parameter Distribution}
\label{sec:impact}

Only the planets whose orbits are oriented within a limited 
range of inclination angles are observed to transit their host stars.
One basic assumption required to make statistical inference from an ensemble of transiting planets is that the orbital inclinations of planets should be distributed randomly with respect to the observer. Following this assumption, 
the impact parameters $b$, which are the minimum planet-star 
projection separations normalized by the radii of the stars during 
the transits, are distributed uniformly for the observed transits. Then
from the observed transits, one may correct for the 
selection effects due to such geometric conditions (``geometric bias'')
to take the number of non-transiting planets into account. For 
circular orbits, the geometric bias is $g_p = R_*/a_p$ for transits with
$0 \leq b \leq 1$.

\begin{figure*}
\includegraphics[scale=0.8]{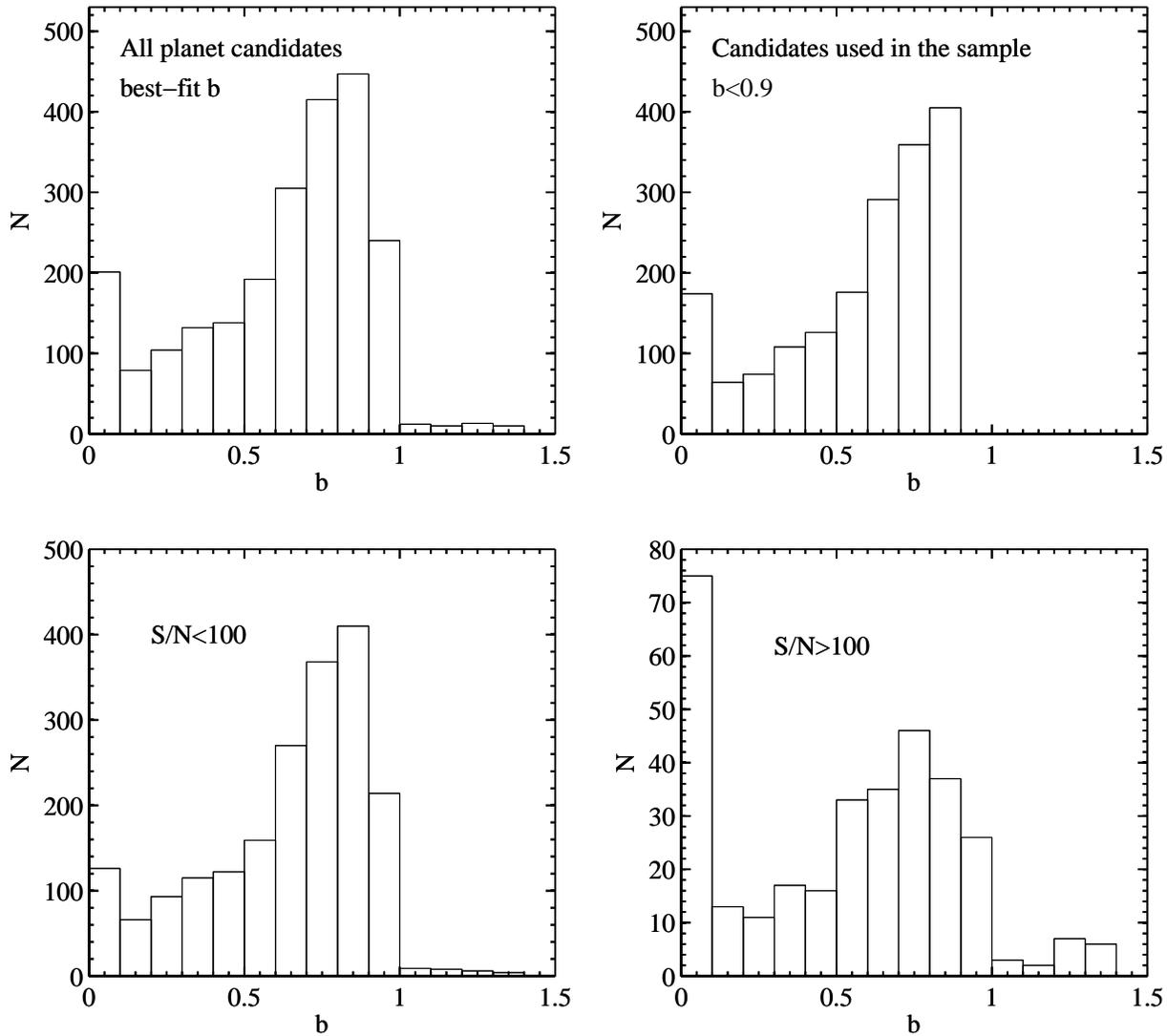}
\caption{Histogram of best-fit impact parameter ($b$) values for $\kepler$
planet candidates reported by B12 (the upper left panel) and the sample
used to derive planet statistics with a upper threshold of $b<0.9$ (the upper right panel). 
Clearly, the reported $b$ is highly skewed toward high values ($\sim 1$), especially for the candidates with lower S/Ns 
(the bottom left panel). This is a very unphysical distribution (see 
discussions in \S~\ref{sec:impact}). 
We also divide the sample into those with lower $(<100)$ and higher 
$(>100)$ S/N in the bottom panels.
For candidates with higher S/N, 
it is less skewed toward $sim 1$ but with a peak at 0. This is 
understandable as at low impact parameter, the transit profile
is hard to distinguish from those at $0$ and the fitting algorithm
may assign $b=0$ as the best fit.
\label{fig:b}}
\end{figure*}

The histogram of best-fit $b$ values for \kepler planets reported
by B12 is plotted in the upper left panel of Figure 2 as well as 
the posterior probability distribution considering Gaussian errors in
the upper right panel (in the latter case, the unphysical values 
for $b<0$ due to the Gaussian distribution are shown). 
The $b$ distribution is far from being uniform, and
it is highly skewed toward large values ($b\sim 0.8-0.9$). 
This unphysical distribution cannot be explained by 
selection effects due to observation thresholds (transits with low $b$
are easier to detect than those with $b\sim 1$ as the former
generally have higher S/N). Note that for candidates with high  
S/N $(>100)$, the distribution is less skewed toward $\sim 1$ but 
with a peak at 0 (see the bottom right panel of Fig.\ref{fig:b}). This is understandable as at low impact 
parameter, the transit profile is hard to distinguish from those 
at $0$ and the fitting algorithm may set $b=0$ as the best fit.

One possible source of the unphysical distribution of 
$b$ skewing toward $\sim 1$ may 
be artifacts or biases introduced by the fitting 
procedures employed by the \kepler team. One possibility is
failure to account for the integration time of the exposure time
in the modeling (\citealt{kipping},  J. Lloyd, B. Gaudi, 
private communications). The other possible source may be that
some of the high-$b$ planets are false positives. Note that B12 also includes a small number of grazing transits with $b$ 
significantly larger than 1, and these candidates are unlikely to be 
of planetary origin. We exclude candidates with impact parameter
larger than $0.9$ from our analysis.

Resolving this discrepancy is beyond the scope of this work. 
In the following analysis, we test whether the planet samples  
with $b<0.6$ and $b<0.9$ result in different distribution functions.
Obviously, given the skewed $b$ distribution, the normalization 
of planet frequency has considerable difference between the 
two samples. We focus on understanding whether the shape of 
the distribution function is affected by the upper threshold of the 
impact parameter $b_\thres$.

\section{Planet Detection Efficiency of \kepler from 
Detection Thresholds}

Besides the geometric selection effect discussed above, the other 
main selection effect is survey selection, which denotes 
an incompleteness due to the detection thresholds of the survey. 
A transit candidate is considered to be detected if (1) the number 
of transit occurrence $\ntr$ exceeds a threshold  and (2) the total 
S/N of the transit signals is greater than the threshold $(\sn)_\thres$. 
We discuss both detection thresholds in detail in the following sub-sections. 

To characterize the survey selection effects, we introduce the planet detection
efficiency $\epsilon(P,R_p)$, which is the fraction of stars 
in the stellar sample for which a planet with period $P$ and 
radius $R_p$ can be detected (i.e., the above two thresholds are 
satisfied). For each star $i$ in a sample with a total of $N_*$ 
stars, the noise $\sigma_i$ and time window during which it is 
observed, $T_{\rm w, i}$, are known. For a hypothetical planet with 
$P$ and $R_{p}$ orbiting this star $i$, we calculate 
$N_{\rm tra}$ and $\sn$ for $100$ uniformly distributed 
phases for the planet transits within time window $T_{\rm w, i}$.
Then among the $100$ simulations, we count how many of them
have both the $N_{\rm tra,\thres}$ and $(\sn)_\thres$ criteria satisfied
to obtain the fraction of phases $\eta_{i}$ where the transits satisfy
the detection criteria. Finally, we obtain 
the detection efficiency for the planet in the sample by summing 
$\eta_{i}$ for all the stars, 
to be $\epsilon(P,R_{p}) = \Sigma_{i=1}^{N_*} \eta_{i}/N_*$.

The intrinsic planet frequency $f_p$ is
defined as,
\begin{equation}
f_p = \frac{{\rm d}^2 N_{p}/N_*}{{\rm d}\log_{10} P{\rm d}\log_{10} R_p},
\label{eq:deffp}
\end{equation}
where $N_{p}$ is the intrinsic number of planets around $N_*$
host stars.
 With both detection efficiency $\epsilon(P,R_{p})$ and 
geometric bias $g_{p}$ known, the intrinsic planet frequency 
$f_p$ can be derived using the relation,
\begin{equation}
\frac{{\rm d}^2 N_{p,\rm det}}{{\rm d}\log_{10} P{\rm d}\log_{10} R_p} 
= N_* f_p \epsilon(P,R_p) g_p,
\label{eq:eqfp}
\end{equation}
where $N_{p,\rm det}$ is the number of planets that pass 
the detection thresholds.

In the following two subsection, we will describe how we calculate the two survey selection criteria:
(1) $\ntr$, and (2) the S/N threshold
\subsection{$\ntr$ Threshold}
\label{sec:window}

We include the effects of the transit window function, which is
important for statistics of long-period planets \citep{gaudiwindow}. 
Out of 122328
stars we have selected, $\sim 68.5\%$ have data over all six quarters,
$\sim 13\%, 1.5\%, 12.4\%, 4\%$ and $0.7\%$ miss $1,2,3,4$ and $5$
quarters, respectively. Over all 6 quarters, the gaps between
quarters and the artifacts amount to a total of $51.5 {\rm d}$,
which is $\sim 10.4\%$ of the duration from the start of
Q1 to the end of Q6 (see Figure 3. for an example that
demonstrates the effect of gaps and Table 1 for a list of the gaps). 
B12 used the Transiting Planet Search (TPS) module
\citep{tps} as the primary algorithm to search for periodic square 
pulses within Q1-Q6 and then sought confirmations in Q7-Q8. 
Strictly speaking, the TPS module finds transit with at least three
occurrences \citep{tps}, but B12 include planet candidates with fewer 
transits occurring in their sample. Moreover, the independent 
searches by \citet{huang} and \citet{corot} over Q1-Q6 
that include transits with less than 3 occurrences only yield $10\%$
more candidates with no obvious preference for long-period ones.
For a detection, we adopt a transit occurrence criterion that at 
least 2 transit occurrences in Q1-Q6 so that it is periodic 
in this window and 3 transit occurrences in Q1-Q8 so that the 
detection is secure. We also vary this criterion to demand 
3 transit occurrences in Q1-Q6 to check whether we obtain 
consistent planet statistics in \S~\ref{sec:disc}. 

In order to evaluate the effect of window functions, for each 
trial period, we make 100 simulations with the center of the 
transits occurring at different times, which are evenly distributed 
within the period. Then we record the number of transit occurrences 
for each quarter in each simulation.
In Fig.\ref{fig:ntran}, we show $f_{\rm Window}$, the fraction of
simulated transits that satisfy the transit occurence criterion
as a function of period. The black line represents a star that has
been observed over all 8 quarters. $f_{\rm Window}$ starts to decrease
from $100\%$ at $P \sim 100 \da$ to $50\%$ at $P \sim 250\da$ then
to $0$ at above $\sim 340 \da$. We also show an example that has one
quarter (Q5) is missing in red line, for which $f_{\rm Window}$ is
typically $\sim 10-20 \%$ smaller at long periods and no transit
satifies the occurrence criterion at $P \gtrsim 300 \da$. This
emphasizes the importance of considering various transit phases for deriving
the frequency of planets with long period beyond 100 days. 

\subsection{S/N Threshold with Box-like Profile}
\label{sec:box}
The statistics of the \kepler planet frequency presented in this
work are completed by 1) using a simple box-like transit profile for both real
and hypothetical planets, and
2) modeling the planet detection threshold with a lower limit in transit
signal-to-noise ratio $\sn > \snth$. This is the same assumption made  by H12 and B11. The simple box-like transit
profile is characterized only by the depth $\delta$ of the transit and 
the transit duration $\tdur$
with the photometric error $\sigma$ for $\tdur$. 
For each star, we have calculated $\sigma$ in each individual 
quarter separately by interpolating the 
published CDPP values (by the \kepler team) at  3$hr$, 6$hr$, 12$hr$ intervals to the desired transit
duration time $\tdur$ (for a 
description of CDPP see \citealt{CDPP}; the CDPP tables
can be downloaded from the official \kepler MAST site).
The total S/N from observing $\ntr$ box-like transits is,
\begin{equation}
\sn = \sqrt{\sum_{k}^{\ntr}\frac{\delta^2}{\sigma_k^2}}.
\end{equation}
The box-like transit profile applies in the limit where the planet-to-star 
radius ratio is small ($R_p \ll R_*$), there is a zero impact parameter ($b=0$), 
and a uniform host star surface brightness profile (no limb-darkening). 
In this limit, $\delta = ({R_p}/{R_*})^2$, and $\tdur=R_*P/(\pi a)$ for circular
orbit. The {$\rm S/N$}s for the candidates
are calculated using the measured transit durations. Both real and hypothetical planets are considered to be detected when $\sn(R_p, P) > (\sn)_{\rm thres}$.

In this limit, the dependency of S/N on the impact parameter $b$
is ignored. In the experiments we carry out below where 
we vary the upper threshold $b_\thres$ for the selection 
of the planet sample, we simply modify the geometric bias
to be $g_p \times b_\thres$. In \citet{dongzhu}, we introduce 
a full framework that takes the effects of limb-darkening and 
ingress/egress into account. In that case, $b_\thres$ also 
introduces changes in the detection efficiency $\epsilon$ since
the S/N detection threshold depends on $b$. Similar to 
\citet{ogle}, we find that adding limb-darkening and 
ingress/egress makes little difference in the inferred distribution.

\section{Results}
\subsection{\kepler Planet Frequency}
\label{sec:frequency}

We first carry out the detection efficiency calculations described 
above for a dense $100 \times 40$ grid of $(P, R_p)$ with $P$ from
$0.3d$ to $500d$ and $R_{p}$ from $0.5R_{E}$ to $32R_{E}$. The grids
are divided uniformly in log space for both $P$ and $R_{p}$.
In the main calculation, we choose 
\footnote{Note that the S/N values we calculate above using CDPP 
are very close to the Multiple Event Statistics (MES) values 
reported by B12, which are the quantities used by the main \kepler 
transit search algorithm TPS which resembles the transit S/N for a 
periodic square-pulse search. MES must be greater than 
$7.1$ in the search conducted by \kepler. We adopt 
a higher threshold $8$, which corresponds to the turnover of 
the right-hand panel of Fig. 7 of \citet{tps}. The S/N of 
the transit fit reported by B12 does not have the cut of 
$7.1$ (with minimum of $4$) and is on average factor of $\sim 2$ 
higher than MES with large variance in ratio between the two 
quantities. Throughout the paper, we use the S/N values calculated 
using CDPP to closely mimic the transit detection processes 
employed by TPS.} $(\sn)_\thres = 8$, 
$N_{\rm tra,\thres} ({\rm Q1-\rm Q6}) = 2$ and $N_{\rm tra,\thres} ({\rm Q1- \rm Q8}) = 3$ and
$b_\thres = 0.9$. All the thresholds are varied in \S~\ref{sec:disc}
to make consistency checks. 
The resulting detection efficiency $\epsilon$, 
and $N_{*,\rm eff} = N_*\times \epsilon \times g_p$, 
which represents the planet sensitivity considering 
both detection efficiency and geometric bias, are shown in 
the left and right panels of Fig. \ref{fig:Incomplete}, respectively.
Beyond $100\days$, $Kepler$'s sensitivity to detect
$R_\earth$ planets drops abruptly.

Then, we divide the $P$ and $R_p$ plane into $15 \times 4$
bins which are uniformly distributed in $\log_{10} P$, 
$\log_{10} R_p$ with $P$ from $0.75d$ to $250d$ and $R_{p}$ 
from $1 R_{E}$ to $16R_{E}$ (see Figure \ref{fig:2d}). In each bin, 
we take the detection efficiency as well as geometric
bias into account and calculate $f_p (P, R_p)$ as 
defined in Equation (\ref{eq:eqfp}) and its $1-\sigma$ uncertainty 
assuming a Poisson distribution. In each bin, $f_p$ is assumed 
to be distributed uniformly in $\log_{10} P$ and $\log_{10} R_p$. For a  bin in which there is no 
planet detected, we compute an upper limit at $90 \%$ confidence 
level. There are 2486 planet candidates in total including B12, 
\citet{huang}, and \citet{corot}, our stellar parameter cuts
limit the number of planets to 1801, and 1347 of these survive 
our detection threshold cut. We examine the effects of adding candidates from \citet{huang} and \citet{corot} and find that excluding these candidates has negligible impact on the derived planet distributions. The bins in the lower right 
corners have the least secure statistics due to low 
 sensitivity in detecting planets and relatively large gradients in 
 the sensitivity. The sensitivity $N_{*,\rm eff} = 
 N_*\times \epsilon \times g_p$ is plotted in red lines in 
 Figure \ref{fig:2d}. 

The intrinsic number of planets ($N_p$ per star) within each period and planet radius bin is shown in  Fig. \ref{fig:period}. The planet radius bins are 
$8-16 R_\earth$ (``Jupiter-size''), $4-8 R_\earth$ (``Neptune-size''),
$2-4 R_\earth$ (``Super-Earth-size'') and $1-2 R_\earth$ 
(``Earth-size''). The bin size in $\log R_p$ (0.3 dex) is chosen
to be larger than the averaged dispersion in $log R_p$ ($0.15$dex) 
due to the uncertainty in KIC $\log g$ estimates. The above-mentioned bins with the least secure statistics are plotted with dash-dotted lines. These 
include the four longest period bins for Earth-size planets ($1 R_\earth<R< 2 R_\earth$) and the longest period bin for  
Super-Earth-size planets ($2 R_\earth<R< 4 R_\earth$).
 
We confirm the the sharp drop below $10\days$ in 
planet frequency 
identified by \citet{howard}. Beyond $10\days$, 
the most striking feature is that the frequency of Neptune-size
planets rises sharply while the smaller planets with $R_p$
from 1-4 $R_\earth$ have frequency consistent with being flat in 
$\log P$. Quantitatively, the frequency of Neptune-size planets 
increases by a factor of $\sim 5$ from $10\days$ to $250\days$. In 
contrast, the frequencies of Earth-size and super-Earth-size planets 
are consistent with flat distributions in $logP$ within 1-2 $\sigma$  beyond $10\days$. The frequency of Jupiter-size 
planets increases more slowly compared to the rise of the 
Neptune-size planets. These trends survive by
varying several observational cuts (discussed in \S~\ref{sec:cuts}) 
so they appear to be robust.

Next we show the cumulative planet frequency for planets
with different sizes in Figure \ref{fig:period_accu}. 
Within $250 \days$, Earth-size and Super-Earth-size
planets have almost the same cumulative frequency $\sim 30 \%$, 
which is $\sim$ 4 times larger than the Neptune-size planet 
frequency ($\sim 7\%$), or $\sim$10 times larger than the Jupiter-size planet 
frequency ($\sim 2.5\%$). The total frequency for all the planets from 1-16$R_{E}$
within 250 $\days$ is $\sim$ 60$\%$. However, the 
absolute normalization is likely not robust as it can vary by a 
factor as large as $\sim 1.5$ depending on various cuts (in 
particular the impact parameter cut) as discussed 
in \S~\ref{sec:cuts} below.

We then show the planet frequency as a function of  planet size 
within three period bins 
(0.4-10, 10-50, 50-250 $\days$) in Fig.\ref{fig:radius}. 
There appear to be clear evolution of planet size 
distribution as a function of period.
At all periods, the dominating population in number is the 
planets with small radii ($R_p < 4R_\earth$). 
There are clear breaks in the distribution function 
at $\sim 3 R_{\earth}$ and $\sim 10 R_{\earth}$. 
At the shortest period 
($<10\days$), below $3 R_\earth$, the planet frequency in $\log_{10} R_p$ increases slowly toward small radii. 
After a relatively steep drop in frequency at $3-4 R_{\earth}$, larger 
planets are consistent with a flat distribution up 
to $\sim 12 R_\earth$. At longer periods ($>10\days$), 
below $3-4 R_\earth$, 
the distribution is consistent with being flat in $\log_{10} R_p$ (or
even consistent with slightly decreasing toward small radii for the 
$P=10-50\,\days$ bin). We caution that planet statistics presented
here are the least secure for $1-2 R_\earth$ at $P\gtrsim 50\,\days$.
Within $P=10-50\days$, planet frequency in $\log_{10} R_p$ 
for planets larger than $\sim 3 R_\earth$ clearly decreases 
for increasing radius up to $\sim 10 R_\earth$.
In the bin with longest periods ($P=50-250\,\days$), 
for planets with $R_\earth = \sim 3-10 R_\earth$, the frequency 
distribution is nearly flat in $\log_{10} R_p$ up 
to $\sim 10 R_\earth$ then it drops sharply at $>10 R_\earth$. 
Overall, at longer period, the relative frequency for big planets
($3 R_\earth<R<10 R_\earth$) compared to small 
planets ($1 R_\earth<R<3 R_\earth$) becomes higher. 

The method presented in this section has the advantage of 
making no assumption on the functional form of planet distribution, 
but the data are binned, which has the implicit
assumption that planet are distributed uniformly within the bins.
Thus, the results may depend on the bin size. We have tested 
the effects of bin sizes by using bins that are factor of 3 smaller, 
and the resulting trends in frequency are consistent with 
those presented above.

\subsection{The maximum likelihood method}

Motivated by the linear trends seen in the log-log plots in 
the period distribution for $P>10\days$ discussed in the previous 
section, we model these trends with power-law dependencies
in period using the maximum likelihood method. This approach 
has the advantage of requiring no binning.

We follow \citet{tt02} and \citet{youdin} to 
calculate the log likelihood function as
\begin{equation}
{\rm ln}(L)=\sum_{j} {\rm ln} (\epsilon_{j}g_{p,j}f_{p})-N_{exp}\,.
\end{equation}
where the sum is taken over all the planet candidates. $\epsilon_{j}$ and $g_{p,j}$ are the 
detection efficiency and the geometric bias as defined above. 
The intrinsic planet frequency $f_{p}$ is defined 
in Eq \ref{eq:deffp} and the assumed analytical form is
\begin{equation}
f_{p}=C\times\left(P/10 {\rm days}\right)^{\beta}\,\,\,\,\,\,{\rm when\,\,\, P>10\days}
\end{equation}
where $\beta$ is also the slope of the intrinsic frequency in the log-log plot.
$N_{exp}$ is the expected number of planets with the assumed $f_{p}$
\begin{equation}
N_{exp}=\int N_{*}\epsilon_{p} g_{p} f_{p}d{\rm log_{10}}Pd{\rm log_{10}}R_{p}\,.
\end{equation}

We numerically solve the maximum log likelihood for
planets in each radius bin (1-2, 2-4, 4-8, 8-16 $R_{\earth}$). The resulting $C$ and $\beta$ are given in table 2. Multiplying $f_p$ with the bin size as in \S 4.1, we derive the planet frequency, which is over plotted in the left panel of Figure \ref{fig:period} as the gray dashed lines. 
Our maximum likelihood fits are consistent with 
the trends in distribution functions described in \S 4.1, confirming our claims that
planets at 1-4 $R_{\earth}$ have a nearly flat distribution in ${\rm log_{10}}$$P$ beyond 10 days, while planets at 4-8 $R_{\earth}$ display a fast increasing distribution in $\log_{10} P$ for increasing period $\propto P^{0.7\pm0.1}$.

We assume power-law distributions with respect to planet period for planets in four different radii bins.
Figure \ref{fig:radius} suggests that the planet
radius distribution function is more complicated than 
simple power-law or broken power-law distribution. We therefore
do not attempt to fit analytical functions to the radii distribution 
with maximum likelihood method. Figure \ref{fig:radius} itself is more instructive than such a 
multi-parameter representation.

\section{Discussion}
\label{sec:disc}
\subsection{Varying Sample Selection Cuts}
\label{sec:cuts}
We vary several sample selection cuts to test the robustness of the derived planet frequency.

First we vary the detection thresholds: $\ntr$, and the S/N threshold.
A S/N threshold=12 (8 is used for the main results) is applied, and the results are shown in the upper left
panel of Fig. \ref{fig:multi}. Obviously smaller planets are 
more affected by making this new cut, and as a results, the 
statistical uncertainty for Earth-size planets becomes much larger.
Nevertheless, the power-law index $\beta$ in period 
distribution is in good agreement with the main results with 
a lower S/N threshold. We also test the case if $\ntr$ 
requires three transits from Q1-Q6, and the results are shown in the upper left panel of Fig. \ref{fig:multi}.
This cut limits the number of planets at the longest period bin. 
Again, $\beta$ is consistent with the main results.
 Next we only choose the bright stars (\kepler magnitude $m_K\le$14.5) in our stellar sample.
These stars on average have less noise than the main sample, thus the transits for small planets have 
higher S/N ratios. The results are consistent with the main ones
for $\beta$.

Given the concern over the skewed $b$ distribution of $Kepler$ planet candidates as discussed in \S 2.2, 
we test the planet frequency with a planet sample having $b<0.6$. This cut causes
bigger changes than those in all previous tests. First, it leads to lower planet frequencies 
($\sim  60\%$ relative to our fiducial case) since this cut
decreases the number of planets by a factor of three while it should only decrease the planet sample
by a factor of 1/0.6=1.7 if the $b$ distribution were uniform. Second, it alters the shape of the distribution 
for small planets and at long period. For planets in both 1-2 $R_{\earth}$ and 2-4 $R_{\earth}$ bins, the power
law index $\beta$ increases compared to the results using $b<0.9$ cut by $1-2 \sigma$. The power-law index for $4-8 R_{\earth}$
is $0.64\pm0.19$, so it is slightly smaller than the main result
but well within uncertainty. See Table 3 for the results of power-law fits using various cuts.

Our conclusion that, beyond 10 days, 
small sized planets (esp. super-earth-size planets) have a nearly 
flat distribution, and Neptune-size planets 
show a fast rising distribution beyond $\sim 10$ days appears to be robust from our various cuts.

\subsection{False Positives \& Blending}
Astrophysical false positives for planet transit candidates usually
involve various scenarios of blending with eclipsing binaries.
Only a small fraction of 
\kepler planet candidates have been confirmed by 
RV (or transit timing variations). It is unlikely that a significant
fraction of \kepler candidates will be confirmed by RV given that most
of them are hosted by relatively dim stars and have masses too low  
to be followed up by RV for existing facilities. Thus 
so far the false positive rates for \kepler candidates are mostly estimated statistically rather than from direct measurements. 
\citet{multiples} 
estimated that $\sim 98\%$ of the planet candidates 
in multi-transiting systems are not due to false positives.
Early statistical estimates on the overall \kepler sample 
according to Galactic models and stellar population synthesis
by \citet{mortonjohnson} claimed that \kepler candidates
have a low rate ($<10\%$) of false positives. However, 
\citet{sophie} found that $\sim 35\%$ of candidates are due to false positives by following up 46 Jupiter-size planet candidates 
with $P<25\days$ from B11 sample. The discrepancy with 
\citet{mortonjohnson} is probably because \citet{mortonjohnson}
did not take M-dwarf eclipsing binaries into account and assumed 
a more stringent vetting procedure than that applied in B11 
(e.g., removing the suspicious V-shape transits, which was not done in B11 but done in B12).
Another possible source of discrepancy is that \citet{mortonjohnson}
assumed a hierarchical triple fraction of $6\%$, but this fraction 
is nearly order-of-magnitude higher for inner binaries with short 
periods \citet{triple}, which is relevant to the close-in giant 
planet candidate sample of \citet{sophie}. Note that these sources 
of discrepancy are most applicable to short-period Jupiter-size
planet candidates, which make up a small fraction of \kepler
planet candidates. The skewed impact parameter distribution 
toward $\sim 1$ discussed in \S~\ref{sec:impact} may also alert us to 
the possibility of false-positive contaminations. In this work, we 
consider a low false-positive rate and do not 
distinguish between planet candidates and planets. Known 
false positives are removed prior to the analysis. Our main 
conclusions on the shape of distribution functions can 
be compromised if there are significant false-positives and the
false-positive rates depend considerably on planet radius and 
period. Systematic efforts in estimating false-positive rates 
such as BLENDER \citep{torres} and \citet{morton2} may help 
to clarify this issue in the future. We also ignore the effects of significant blending in the light curve \citep{seager}. 
The primary effect of blending is to dilute the transit depth, and as 
a result, the planet radius can be underestimated. In 
addition, derived transit parameters such as impact parameter 
can also be altered due to blending.

\subsection{Comparison with Previous Work}
Our approach to computing detection efficiency is similar to 
H12 while our stellar sample
is factor of $\sim 2$ larger than the main sample in H12 and 
the planet sample is factor of $\sim 3$ larger. Importantly, the B12 planet candidates we use are derived from a longer observing span 
(Q1-Q6) than the B11 sample used by H12, and the
improved planet detection algorithm in B12 is likely much more efficient than B11 and probably has a high level of completeness
up to $\sim 250 \days$. We have also considered the effect of 
the observing window function, 
which is essential for studying the statistics of long-period planets. With these improvements, we are able to probe a larger 
parameter space $(R_{p}\ge 1 R_{\earth}$, $P<250 \days)$ compared with H12 ($R_{p}\ge 2 R_{\earth}$, $P<50 \days$). For the overlapping parameter space, our results are consistent with those of H12. 

We may also compare with the frequency of small planets from RV surveys \citep{mayor11}. Detailed comparison 
would require modeling the mass-radius relation, which has a large uncertainty for the majority of \kepler planets of interest. We 
only attempt to make a tentative comparison on the 
broad features and general trends.
\citet{mayor11} found that more than $50\%$ of solar-type stars host ``at least one planet of any mass'' 
within $\sim 100\days$. This is broadly consistent with our results that $50 \%$ of \kepler solar-type stars host planets
with $R_p > 1 R_\earth$ with $P<100\days$. 
\citet{mayor11} have 
also suggested that the frequency of planets with
$M_p\sin i< 30 M_\oplus$ may drop sharply for $P>100$ days, although they caution that this could be an artifact of selection bias 
(see the red histogram of Fig. 14 and the discussions in Sec 4.4
in their paper).
Therefore, it is of interest to determine whether there is evidence for
a parallel drop in planets in the Kepler data. We focus on the 
planet radius bin $2<R_p/R_\oplus<4$, which probably contains 
a large fraction of planets in the mass bin  
considered by \citet{mayor11}.
After correcting for incompleteness, \citet{mayor11}
found that planet frequency drops by factor of $\sim 3.5$ from
the period bin $[56, 100] {\rm days}$ to 
$[100, 160] {\rm days}$. To be specific, we ask how many planets would be
expected in our 100<P<160 bin if the underlying frequency fell by
a factor 3.5 at this boundary.  We find that $\sim 7$ planets would 
be expected while 23 are actually detected, which is not consistent
with Poisson statistics.  Therefore, the available \kepler data
 appear to be in tension with the suggestion of $100\,{\rm days}$ frequency drop by \citet{mayor11}. A future \kepler release would
 be able to definitively test this claim by probing small planets 
 at longer period.

\subsection{Implications}
The planet distribution in period and radius presented 
in this paper may bear the imprints of planet formation, migration, dynamical evolution and possibly other physical processes (e.g, \citealt{idalin, population, kenyon, eva}). B11 and H12 found a sharp 
decline in planet frequency below $\sim 10$ days. Our analysis 
of planets with longer periods reveals that at $P>10$ days,
planets at all sizes appear to follow smooth power-law distributions 
up to $250 \days$: either a nearly flat distribution in 
$\log P$ for small planets ($< 4 R_{\earth}$) or a rising 
distribution for larger planets ($> 4 R_{\earth}$). 
In particular, Neptune-size planets ($R_p = 4-8 R_\earth$)
have significantly increasing frequency with periods from 
$\sim 10$ to $\sim 200$ days. We are not aware of any
formation or migration theories that predict such distributions. 
Planet size distribution 
evolves with period, and generally the relative fractions 
for big planets increase with period, as shown in 
Fig. \ref{fig:radius}. The exception is planets with the largest 
sizes $R_p>10 R_{\earth}$, whose relative fraction drops
sharply at long period $P>50\days$. 
This is consistent with the finding by \citet{inflated}, and may have implications for the radius inflation mechanisms of the Jovian planets.  
Another distinct break is at $\sim 3$ $R_{\earth}$ in planet 
radius distribution at all periods. The $\sim 3 R_{\earth}$ break was
found by \citet{break} and \citet{youdin} for short-period \kepler
planets in B11 and was regarded as evidence for core-accretion formation 
scenarios.

\acknowledgements 
We thank Andy Gould, Boaz Katz and Scott Tremaine for carefully reading the
manuscript and making helpful comments. We are grateful for useful 
discussions with Fred Adams, Scott Gaudi, Lee Hartmann, Chelsea Huang, Jennifer Johnson, David Kipping, James Lloyd, Tim Morton, and Dave Spiegel. S.D. was supported through a Ralph E. and 
Doris M. Hansmann Membership at the IAS and NSF grant AST-0807444. Work by
SD was performed under contract with the California
Institute of Technology (Caltech) funded by NASA
through the Sagan Fellowship Program.
Z.Z. was supported by NSF grant AST-0908269 and Princeton University. 
Z. Z. acknowledges support by NASA through Hubble Fellowship grant HST-HF-51333.01-A awarded by the Space Telescope Science Institute, which is operated by the Association of Universities for Research in Astronomy, Inc., for NASA, under contract NAS 5-26555.

{\it Note added:} During the refereeing process of the manuscript, we
learned an independent study by \citet{fressin} published after
our submission. They used the same release of {\it Kepler} planet candidates
as in our paper to study the false positive rate and frequency of planets
with periods less than 50 days (as compared to $< 250$ days in this 
paper). Their results on planet frequency (their Fig. 7) are in
excellent agreement with those presented in Fig. 9 in this paper,
although the approaches are different in detail. They found generally
low false positive rate for the majority of {\it Kepler} candidates with a
estimated global false positive rate of $\sim 9.4\%$, supporting the assumption
of a low false positive rate adopted in this paper.

\bibliographystyle{apj}

\begin{figure*}
\includegraphics[scale=0.6]{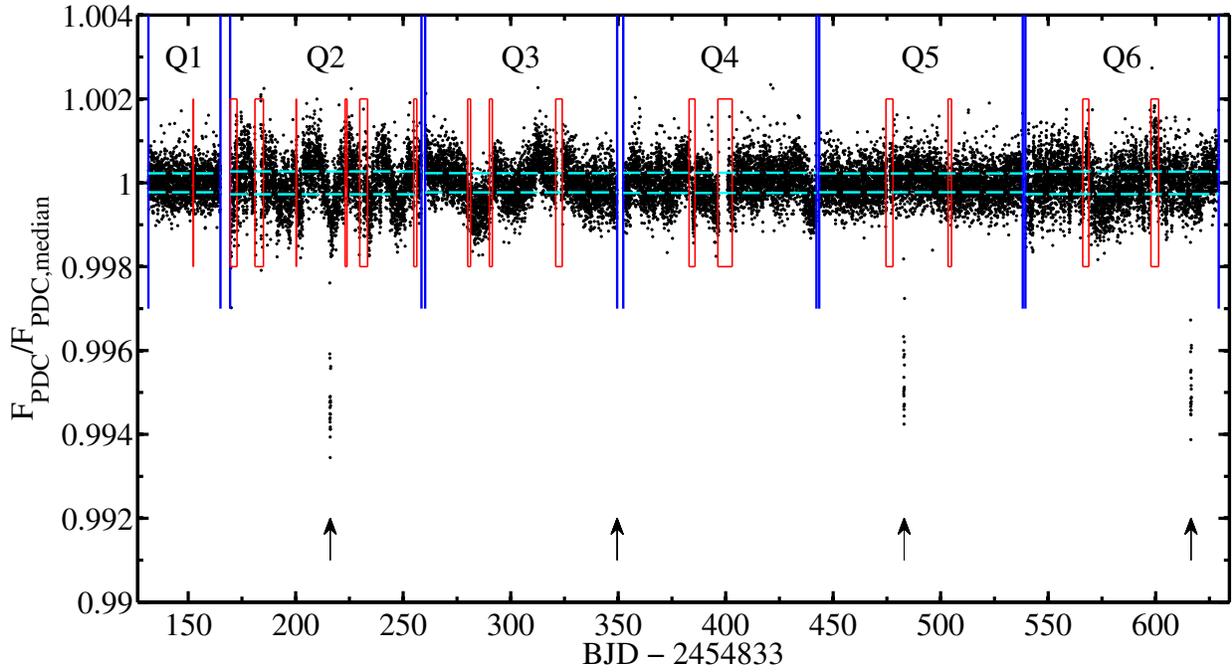}
\caption{Example transit light curve demonstrating the 
importance of the window function for a long-period transit. 
Quarter gaps are between the blue lines, while
other gaps (Table 1) are marked as the red boxes. The arrow indicates the transits,
and one of the transits accidentally falls into the quarter gap
between Q3 and Q4.
}
\end{figure*}

\begin{figure}
\includegraphics[scale=1.0]{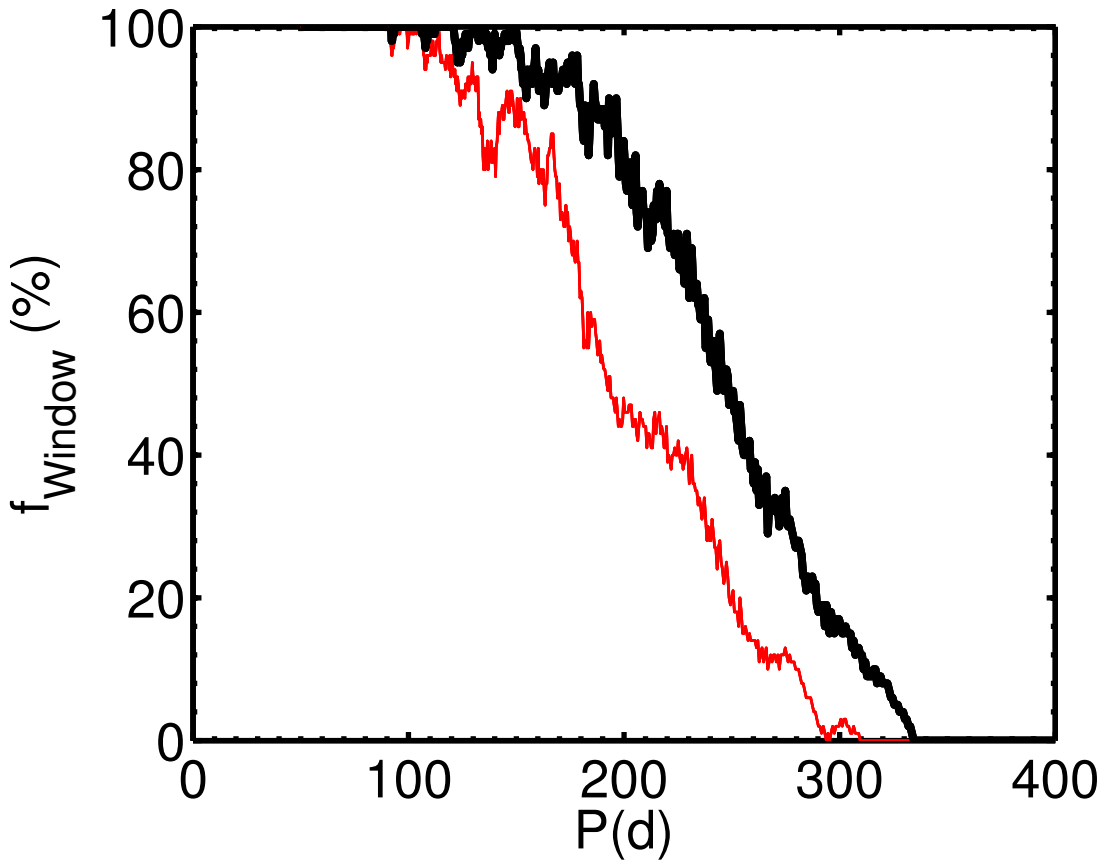}
\caption{Window function $f_{\rm window}$, defined as the fraction of
simulated transits that satisfies the transit occurrence criterion as a function of period.
The black line represent a star that has been observed over all 8 quarters. If
Q5 is missing, $f_{\rm window}$ is plotted as the red curve.  $f_{\rm window}$ is important
for deriving the frequency of planets with long period ($>100$ days)
\label{fig:ntran}}
\end{figure}

\begin{figure}
\includegraphics[width=0.8\textwidth]{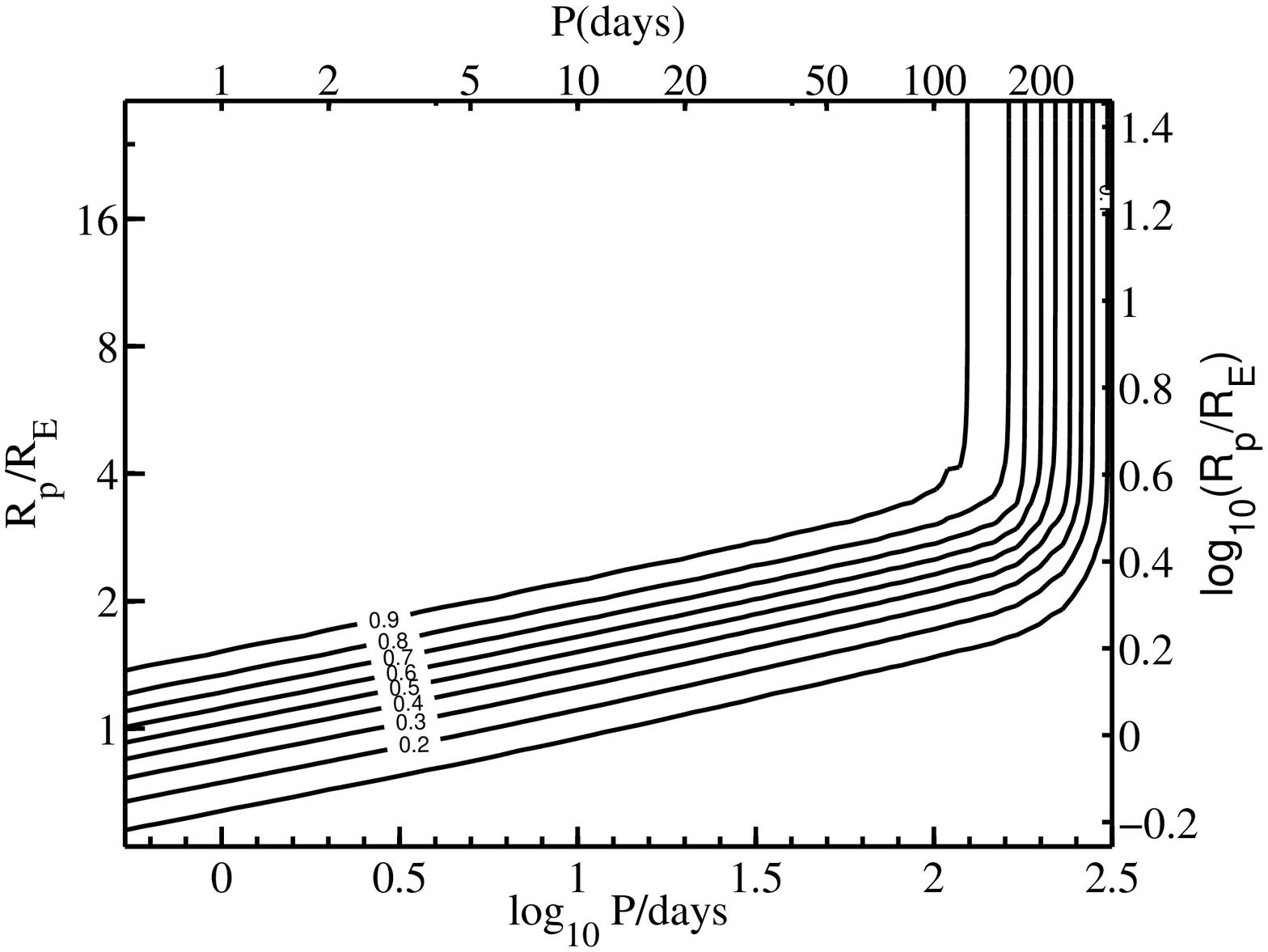} \hfil
\includegraphics[width=0.8\textwidth]{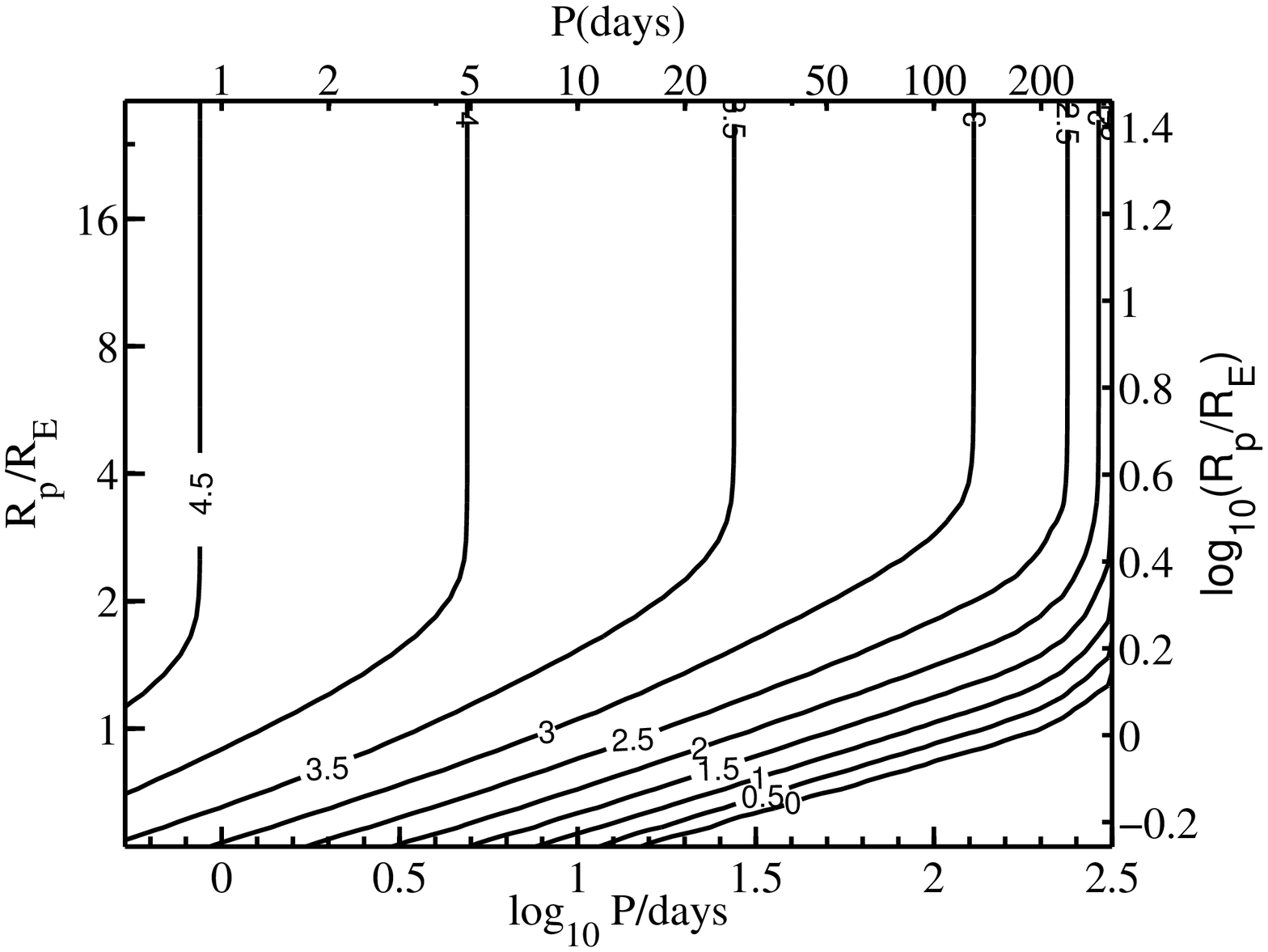}\\
\caption{Upper panel: the derived detection efficiency for
our stellar sample.
Lower panel: the detection sensitivity considering
 the geometric bias and the detection efficiency. The contour
 is shown in log$_{10}$ of 
the number of planets that can be detected if every star in our
sample has a planet at the given $R_{p}$ and $P$. 
 \label{fig:Incomplete}}
\end{figure}

\begin{figure*}
\includegraphics[scale=0.9]{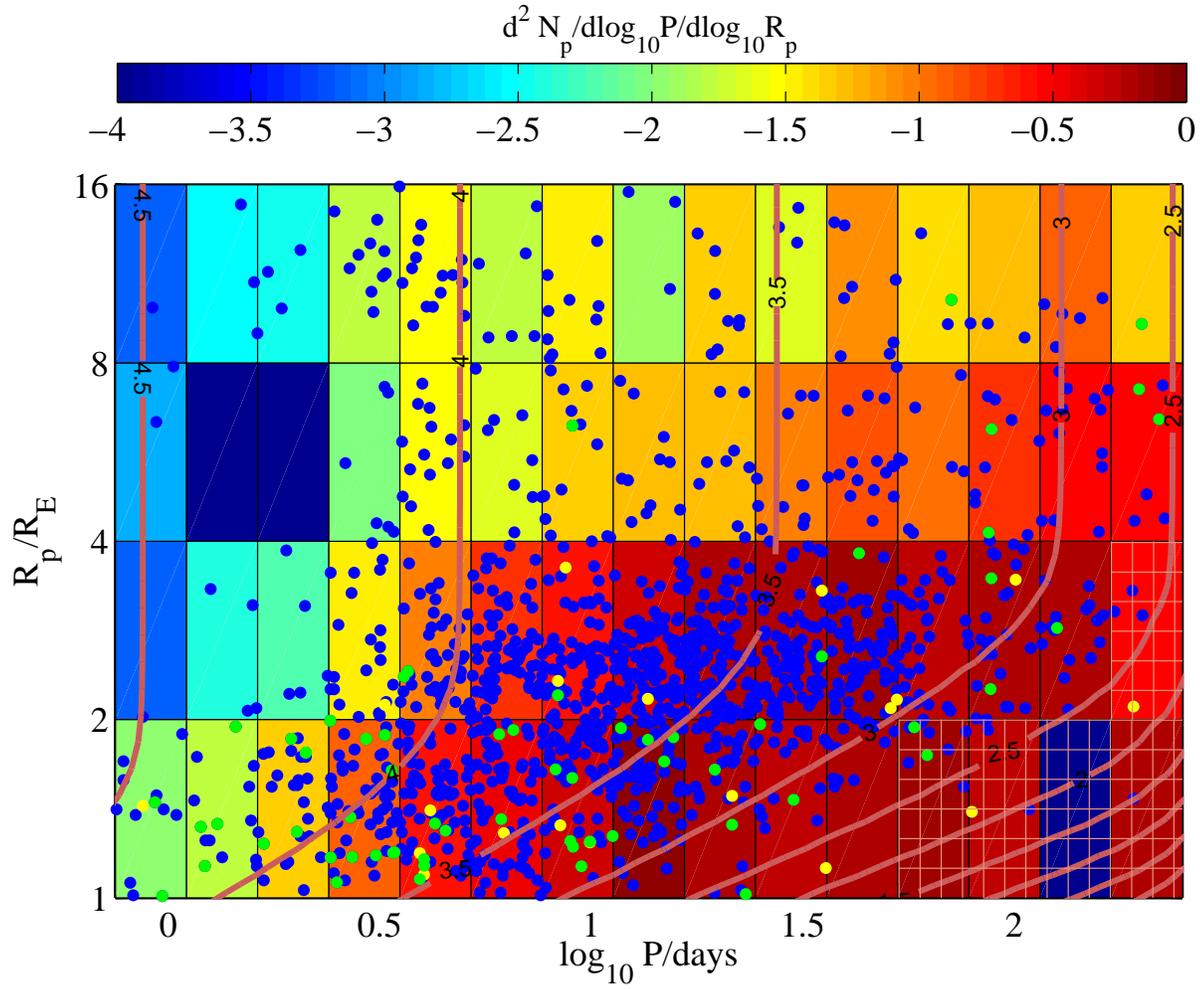}
\caption{Planet frequency as a function $P$ and $R_{p}$. All selected
planet candidates are over plotted including B12 (blue dots), 
\citet{huang} (green dots),
and \citet{corot} (yellow dots). The planet sensitivity shown
in the right panel of Fig.\ref{fig:Incomplete} is also plotted.  The lower right
corner marked with the small grids has the least secure statistics since the sensitivity is low and the gradient in sensitivity is 
relatively large. \label{fig:2d}}
\end{figure*}

\begin{figure*}
\includegraphics[scale=1.00]{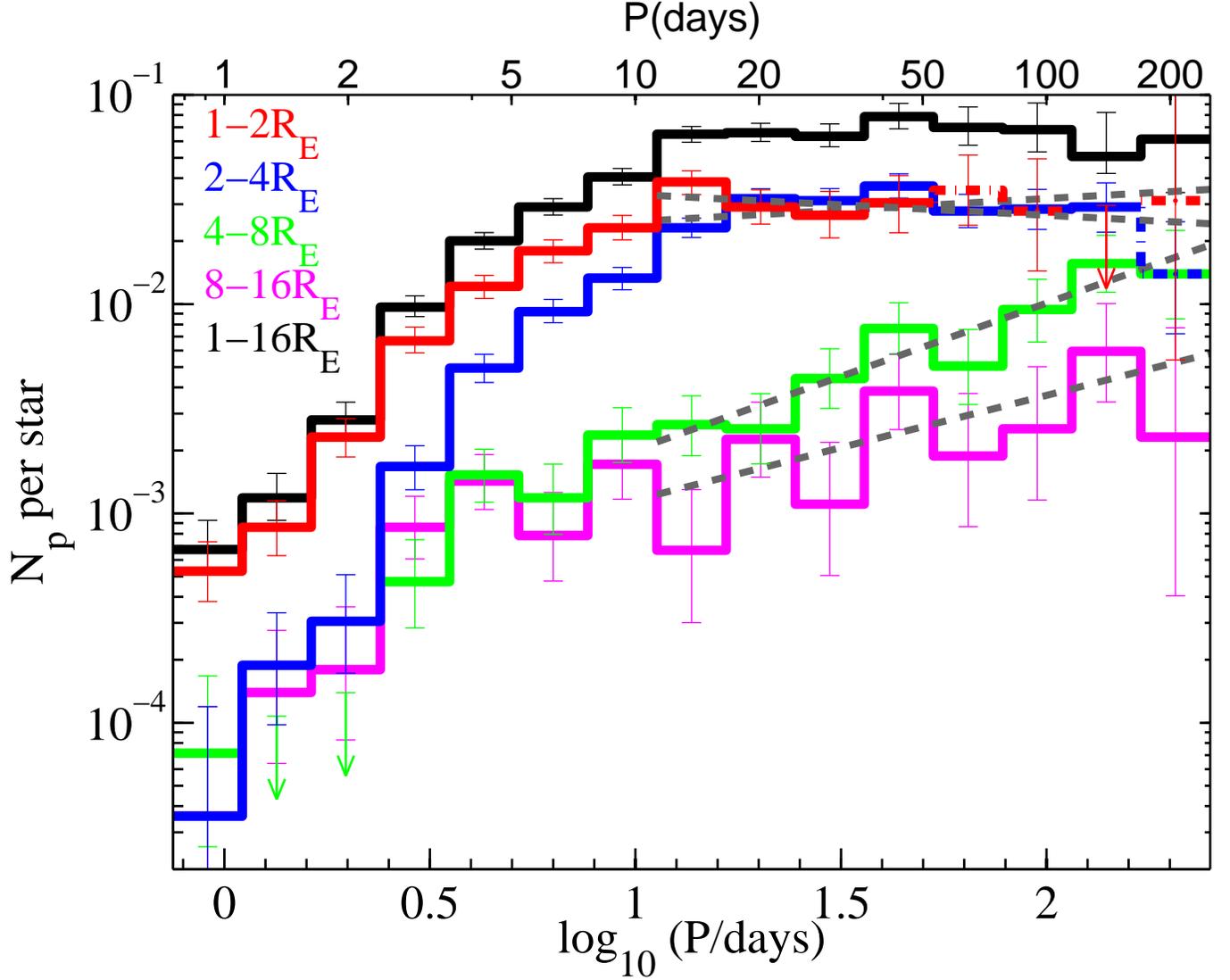} \\
\caption{Intrinsic number of planets per star at different planet radius and period bins plotted as a function of period. 
The histograms with error bars in various colors represents 
different planet-radius bins (red: $1-2 R_\earth$, blue: 
$2-4 R_\earth$, green: $4-8 R_\earth$, magenta: $8-16 R_\earth$, black: $1-16 R_\earth$). The dash-dotted part of the histograms
are for the bins with the least secure statistics, corresponding 
to the bins marked with small grids in Fig.\ref{fig:2d} and the
statistics in those bins are the least trustworthy.
The maximum likelihood best fits in power-law distribution 
as a function of period for planets beyond 10 days at each 
planet radius bin  are over plotted as the  gray dashed lines. 
For orbital period $P>10\da$, 
the planet frequency d$N_p$/d$\log$P
for ``Neptune-size'' planets ($R_p = 4-8 R_\earth$) increases 
with period as $\propto P^{0.7\pm0.1}$. In contrast, 
d$N_p$/d$\log$P for ``super-Earth-Size'' ($2-4 R_\earth$) as well 
as ``Earth-size'' ($1-2 R_\earth$) planets are consistent with a nearly flat distribution as a function of period ($\propto P^{0.11\pm0.05}$
and $\propto P^{-0.10\pm0.12}$, respectively), 
and the normalizations are remarkably similar at 50 d (within a factor of 
$\sim 1.5$). Detailed discussion see \S~\ref{sec:frequency}
\label{fig:period}}
\end{figure*}

\begin{figure*}
\includegraphics[scale=0.9]
{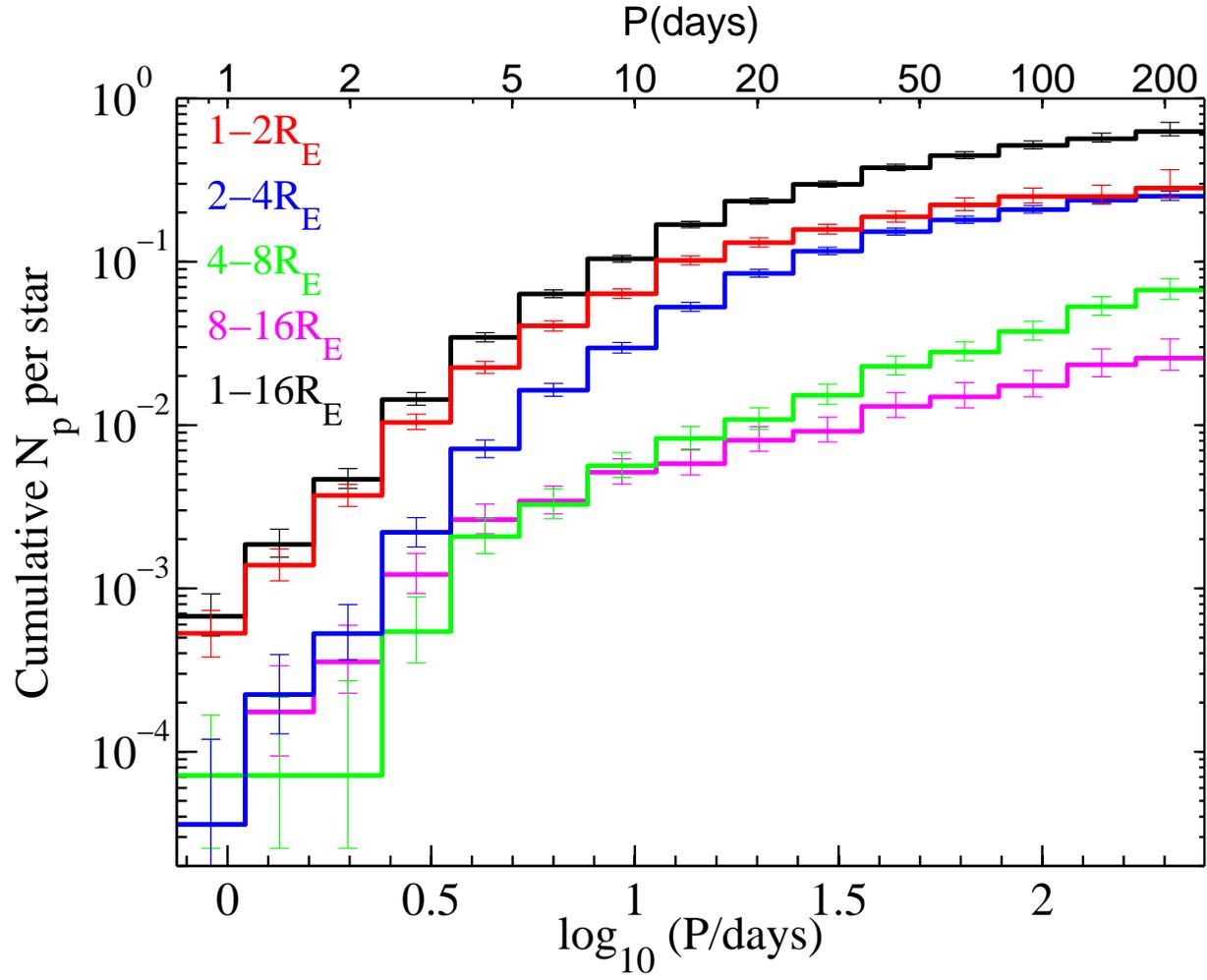} \\
\caption{Cumulative distribution of the intrinsic number of planets per star within $P$ for planets at different radius bins. The color scheme is the same as in Fig \ref{fig:period}.
\label{fig:period_accu}}
\end{figure*}

\begin{figure*}
\includegraphics[scale=0.9]{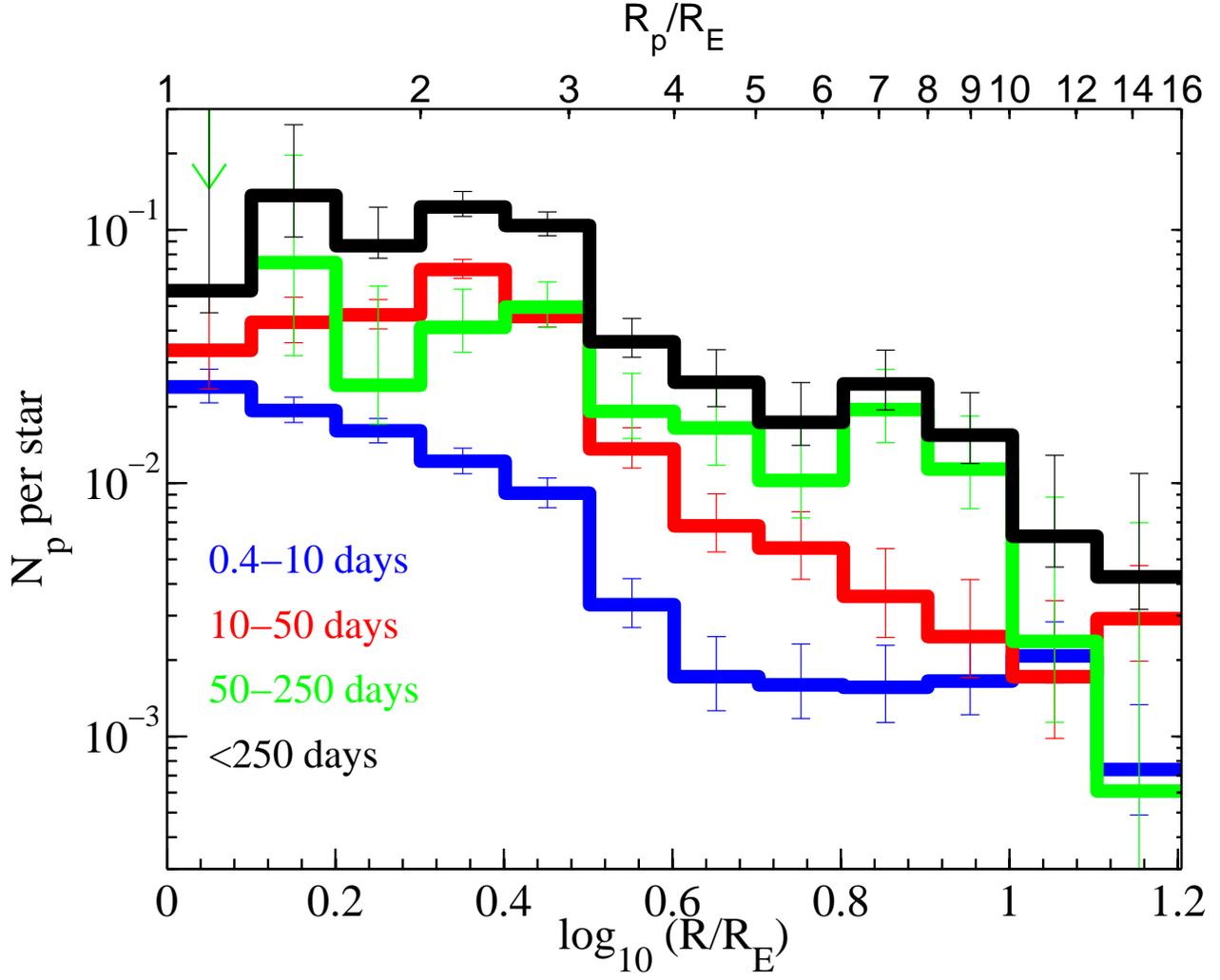}
\caption{Number of planets per star as 
a function of planet radius for planets at different
period bins (blue: $0.4-10 \days$, red: 
$10-50 \days$, green: $50-250 \days$,  black: $<250 \days$). There is considerable evolution in 
size distribution as a function of period. 
There seems to be clear breaks in the size 
distribution functions at 3 and 10 $R_{\earth}$.
The relative fraction of large planets at $3-10 R_{\earth}$ 
compared to small planets ($1-3 R_{\earth}$) increases 
with period. Detailed discussion see \S~\ref{sec:frequency}
\label{fig:radius}}
\end{figure*}

\begin{figure*}
\subfigure[]{
\includegraphics[scale=0.7]{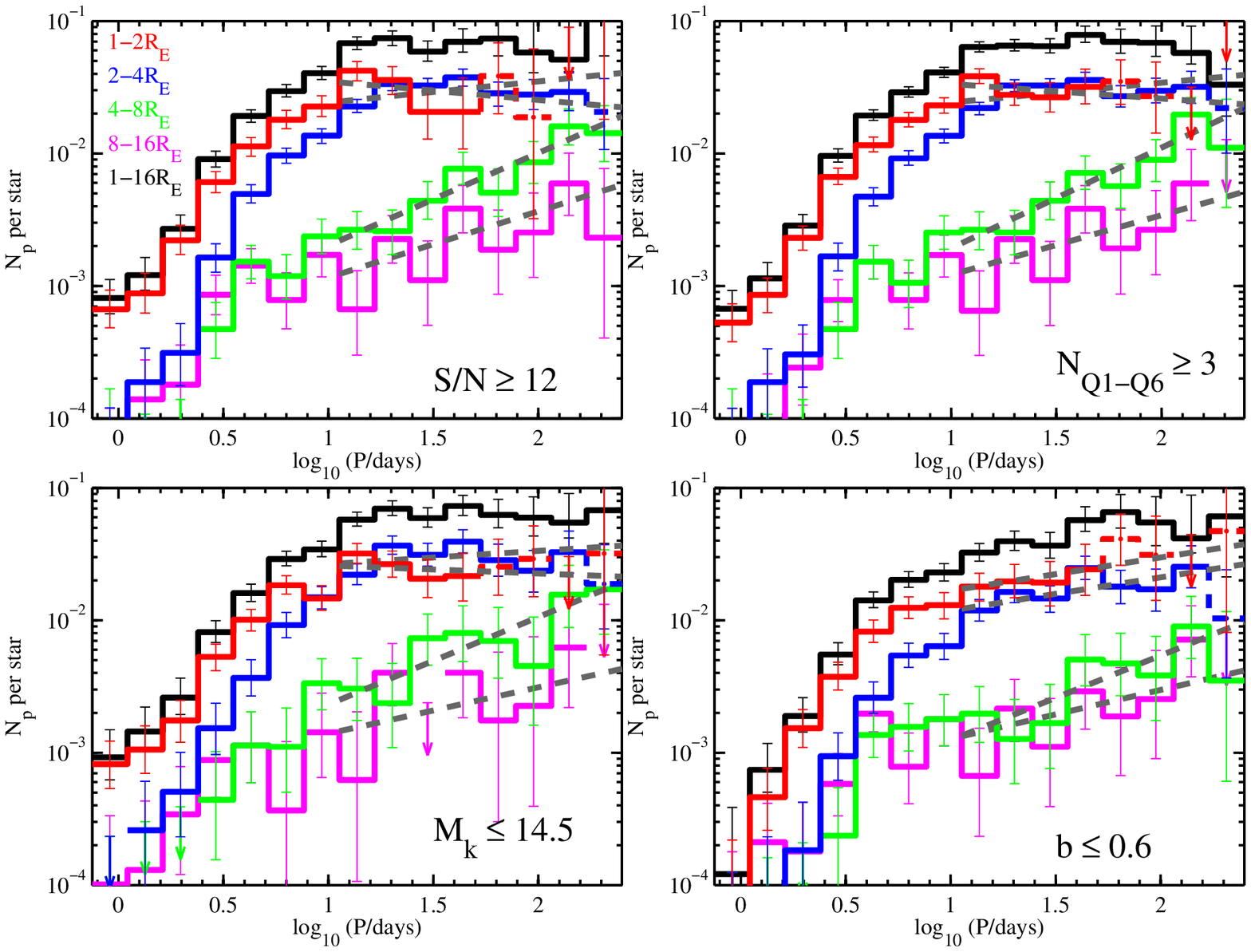}
}
\subfigure[]{
\includegraphics[scale=0.35]{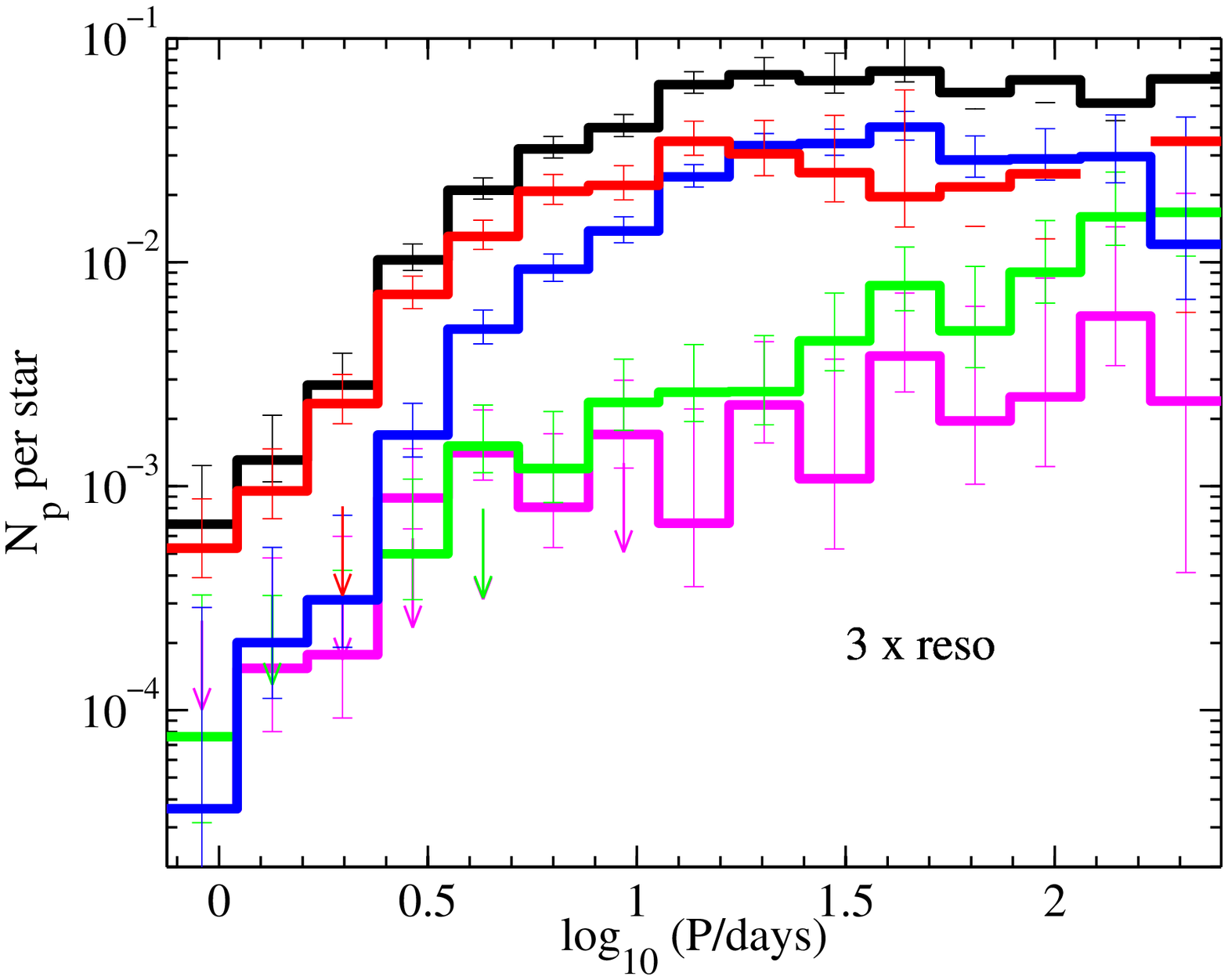}
}
\caption{(a) Results of tests by varying the cuts in sample selections.
The results are presented the same way as in Figure \ref{fig:period}.
Lower left: Cutting stellar sample with Kepler magnitude 
$m_{k}\le$14.5 (rather than 16 for the main analysis).
Lower right: impact parameter cut of $b<0.6$ 
rather than $b<0.9$ for the main analysis.
The upper left and upper right panels: planet detection thresholds cuts (Q1-Q6) transit number larger than 3 rather than 2 on 
the left; S/N$\ge$12 rather than 8 on the right). See \S~\ref{sec:cuts} for discussion. (b) Results by making the bin size in the frequency calculations 3 times smaller. Bin sizes have little effects on the resulting distributions.
\label{fig:multi}}
\end{figure*}

\newpage
\newpage
\begin{table}
\begin{center}
\caption{Gaps in \kepler light curves \label{tab2}}
\begin{tabular}{ccc}

\tableline\tableline
Gap Start  &   Gap end   &  Comments\\
BJD-2454833 & BJD-2454833 & \\
\hline
152.2720 & 152.4740 &  \\
164.9938 &169.5098 &  Gap Q1 \& Q2 \\
169.5195  & 172.7300  & \\
  181.0324  & 185.0000 & \\
  200.1597  & 200.3657 & \\
  222.9826  & 223.8494 & \\
  229.8074  & 233.4153 & \\
  254.8999  & 256.3283 & \\
258.4773 & 260.2141 & Gap  Q2 \& Q3\\
 280.0536  & 281.3308 & \\
  290.0661  & 291.4246 & \\
  320.9617  & 323.9400 & \\
349.5046 & 352.3651 & Gap  Q3 \& Q4\\
  382.9368 & 385.7300 & \\
  396.3515 & 403.0000 & \\
442.2121\tablenotemark{a} & 443.4785 & Gap  Q4 \& Q5\\
  474.5202  & 477.8000 & \\
  503.4133  & 505.0200 & \\
538.1713 & 539.4398 & Gap Q5 \& Q6\\
 566.0423 & 568.9000 & \\
  597.7961 & 601.4000 & \\
  \hline
\end{tabular}
\end{center}
\tablenotetext{1}{For the stars that were only observed part of Q4 due to the malfunction of the CCD, the gap start extends to 373.2282.}
\end{table}

\newpage
\begin{table}
\begin{center}
\caption{Power-law Fits to $\kepler$ Planet Frequency with Periods from 10 days to 250 days with $f_p(P,R_{p})=
C\times\left(P/10 {\rm days}\right)^{\beta}$ \label{tab3}. $f_p$ is defined in Equation (\ref{eq:deffp}).}

\begin{tabular}{ccc}
\tableline\tableline
Planet radii &  C  &  $\beta$\\
$R_{\earth}$ &    &   period power\\
\hline
1-2 & 0.66$\pm$0.08 & -0.10$\pm$0.12\\
2-4 & 0.49$\pm$0.03 &   0.11$\pm$0.05\\
4-8  & 0.040$\pm$0.008  & 0.70$\pm$0.1 \\
 8-16  & 0.023$\pm$0.007 & 0.50$\pm$0.17\\
 \hline
\end{tabular}
\end{center}
\end{table}

\newpage
\begin{table}
\begin{center}
\caption{Similar to Table \ref{tab3} $f(P,R_{p})=C\times\left(P/10 {\rm days}\right)^{\beta}$,
except by varying cuts in sample selection.\label{tab3}.
$f_p$ is defined in Equation (\ref{eq:deffp}).}
\begin{tabular}{ccc}
\tableline\tableline
Planet radii &  C  &  $\beta$\\
$R_{\earth}$ &    &   period power\\
\hline
S/N$\ge$12&&\\
\hline
1-2 & 0.69$\pm$0.10 & -0.14$\pm$0.2\\
2-4 & 0.48$\pm$0.03 &   0.16$\pm$0.06\\
4-8  & 0.040$\pm$0.008  & 0.70$\pm$0.12 \\
 8-16  & 0.023$\pm$0.007 & 0.50$\pm$0.17\\
 \hline
 N$_{Q1-Q6}\ge$3&&\\
\hline
1-2 & 0.66$\pm$0.08 & -0.11$\pm$0.13\\
2-4 & 0.48$\pm$0.03 &   0.15$\pm$0.07\\
4-8  & 0.038$\pm$0.008  & 0.76$\pm$0.13 \\
 8-16  & 0.024$\pm$0.007 & 0.45$\pm$0.2\\
 \hline
 m$_{K}\le$14.5&&\\
\hline
1-2 & 0.51$\pm$0.07 & -0.06$\pm$0.15\\
2-4 & 0.52$\pm$0.06 &   0.10$\pm$0.08\\
4-8  & 0.046$\pm$0.015  & 0.66$\pm$0.19 \\
 8-16  & 0.028$\pm$0.015 & 0.35$\pm$0.31\\
 \hline
b$\le$0.6&&\\
\hline
1-2 & 0.33$\pm$0.06 & 0.25$\pm$0.17\\
2-4 & 0.23$\pm$0.03 &   0.25$\pm$0.1\\
4-8  & 0.025$\pm$0.008  & 0.64$\pm$0.19 \\
8-16  & 0.025$\pm$0.01 & 0.37$\pm$0.23\\
\hline

\end{tabular}
\end{center}
\end{table}

\end{document}